\documentclass[
    ,final            
,numberedheadings 
  ]
  {aipproc}

\layoutstyle{8x11double}

\begin{document}

\title{Future Probes of the Neutron Star Equation of State Using X-ray Bursts}

\author{Tod E. Strohmayer}{
  address={Laboratory for High Energy Astrophysics, NASA's Goddard
Space Flight Center, Greenbelt, MD 20771} }

%
%

\begin{abstract}

Observations with NASA's Rossi X-ray Timing Explorer (RXTE) have
resulted in the discovery of fast (200 - 600 Hz), coherent X-ray
intensity oscillations (hereafter, ``burst oscillations'') during
thermonuclear X-ray bursts from 12 low mass X-ray binaries
(LMXBs). Although many of their detailed properties remain to be fully
understood, it is now beyond doubt that these oscillations result from
spin modulation of the thermonuclear burst flux from the neutron star
surface. Among the new timing phenomena revealed by RXTE the burst
oscillations are perhaps the best understood, in the sense that many
of their properties can be explained in the framework of this
relatively simple model. Because of this, detailed modelling of burst
oscillations can be an extremely powerful probe of neutron star
structure, and thus the equation of state (EOS) of supra-nuclear
density matter. Both the compactness parameter $\beta = GM/c^2 R$, and
the surface velocity, $v_{\rm rot} = \Omega_{\rm spin} R$, are encoded
in the energy-dependent amplitude and shape of the modulation
pulses. The new discoveries have spurred much new theoretical work on
thermonuclear burning and propagation on neutron stars, so that in the
near future it is not unreasonable to think that detailed physical
models of the time dependent flux from burning neutron stars will be
available for comparison with the observed pulse profiles from a
future, large collecting area X-ray timing observatory. In addition,
recent high resolution burst spectroscopy with XMM/Newton suggests the
presence of redshifted absorption lines from the neutron star surface
during bursts. This leads to the possibility of using large area, high
spectral resolution measurements of X-ray bursts as a precise probe of
neutron star structure. In this work I will explore the precision with
which constraints on neutron star structure, and hence the dense
matter EOS, can be made with the implementation of such programs.

\end{abstract}

\maketitle


\section{Introduction}

Thirty five years after their discovery we are still largely in the
dark as to the internal structure of neutron stars. The central
densities of these objects are so high that an understanding of the
equation of state (EOS) of the matter at their cores lies
tantalizingly close yet still beyond the reach of the present
predictive powers of theory. In an astrophysical context this is
reflected in our inability to predict the radius, $R$, of a neutron
star of a given mass, $M$. This quantity, the mass - radius relation,
depends directly on the EOS.  What we would sorely like to do is to
plot the locations of real neutron stars in the mass - radius plane
and in effect measure astrophysically the dense matter EOS. 

There is a direct connection between the dense matter EOS and the
fundamental physics of nucleon interactions. For example, Lattimer \&
Prakash \cite{LM01} have demonstrated that accurate measurements of
neutron star radii provide a determination of the pressure of matter
at nuclear saturation density.  This quantity is in turn directly
related to the nuclear symmetry energy and the isospin dependence of
the nuclear interaction. Moreover, constraints on the maximum mass of
neutron stars would limit the presence of ``exotic'' condensates in
neutron stars as well as bound the central density and thus the
highest densities achievable in nature (see \cite{LM01} and references
therein).

\section{Measuring Masses and Radii}

Precise inferences on the EOS require accurate measurements of neutron
star masses and radii.  For some young, binary neutron star pulsars,
accurate masses have been deduced from relativistic orbital effects
(see Thorsett \& Chakrabarty \cite{TC}). However, little is known
about the radii for this sample.  For the older, accreting neutron
star binaries, there is to date still precious little direct neutron
star mass information. In a few cases where mass constraints can be
attempted there is some evidence for ``massive'' neutron stars, with
masses significantly greater than the canoncial $1.4 M_{\odot}$ (see
Orosz \& Kuulkers \cite{OK99}, for example).

There are a number of different methods by which masses and radii can
be estimated.  Since the space here is inadequate to provide an
exhaustive review of all these I will rather briefly discuss several
areas of recent advancement and explore how the promise of these new
findings can be realized with future, large area X-ray observatories.

\subsection{Burst Oscillations}

Burst oscillations were first discovered as strong, discrete peaks in
Fourier power spectra of X-ray time series accumulated during
thermonuclear X-ray bursts from some neutron star LMXBs (see
Strohmayer et al. \cite{Strohall96}). For an overview of their
observational characteristics see Strohmayer \& Bildsten \cite{SB03}
and the review by Muno elsewhere in this volume. In their discovery
paper Strohmayer et al. \cite{Strohall96} suggested that the burst
oscillations result from spin modulation of the thermal burst flux,
and their is now compelling evidence to support this conclusion. The
initial evidence included; the large modulation amplitudes at the
onset of bursts, the time evolution of the pulsed amplitude during the
rise of bursts (Strohmayer, Zhang \& Swank \cite{szs}), the coherence
of the oscillations (Smith, Morgan \& Bradt \cite{SMB}; Strohmayer \&
Markwardt \cite{sm99}; Muno et al. \cite{2000ApJ...542.1016M}), and
the long term stability of the oscillation frequencies (Strohmayer et
al. \cite{s98b}). In the last few years the observations of highly
coherent, orbitally modulated pulsations in a superburst from 4U
1636-53 (Strohmayer \& Markwardt \cite{sm02}), and burst oscillations
at the known spin frequencies of two accreting millisecond pulsars;
SAX J1808.4-3658 (Chakrabarty et al. \cite{Chaketal03}), and XTE
J1814-338 (Strohmayer et al. \cite{Setal03}) have solidified the spin
modulation paradigm.

In the context of using burst oscillations as probes of neutron stars,
the importance of knowing that they are the result of photon emission
from a non-uniform brightness pattern on the neutron star surface
should not be underestimated. For many astrophysical phenomena a basic
understanding of the emission geometry still remains controversial (an
example is the X-ray emission from stellar mass black holes). The
emission and propagation of photons from the surfaces of rapidly
rotating neutron stars are strongly dependent on relativistic
effects. For example, the amplitude of pulsations is affected by
gravitational light deflection which depends on the compactness,
$\beta$, and the shape (harmonic content) of the pulses is influenced
by the rotational velocity, $v_{{\rm rot}} = 2\pi
\nu_{\rm spin} \sin i$, which depends directly on the stellar radius, $R$, 
and the system inclination, $i$.  A number of studies have attempted
to use these effects to place constraints on the masses and radii of
neutron stars.  Miller \& Lamb \cite{ml98} explored the strength and
harmonic content of pulses from point-like hot spots. They showed that
knowledge of the angular and spectral dependence of the surface
emissivity is important in obtaining accurate constraints. Nath,
Strohmayer \& Swank \cite{nss02} modelled bolometric pulse profiles
observed with the RXTE Proportional Counter Array (PCA) during the
rising portion of X-ray bursts from 4U 1636-53. They concluded that
models with a single hot spot did not yet strongly constrain the mass
and radius. They modelled the emission from a circular, linearly
growing hotspot, and included light deflection in a Schwarzschild
spacetime. Weinberg, Miller \& Lamb \cite{wml01} have explored the
pulse profiles produced by rotating neutron stars, including the
rotational Doppler shifts and aberration of the emissivity. They
concluded that pulse profile fitting is to be preferred over other
indirect measures of the harmonic content, such as Fourier amplitudes.
Muno, Ozel, \& Charkrabarty \cite{moc02} have explored the amplitude
evolution and harmonic content of cooling phase burst oscillations
from a number of different sources.  They used the observed limits on
harmonic content to constrain the location and size of the hot spot or
spots responsible for the observed modulations.

An observational limitation of these attempts has been the inability
to detect any harmonic signals in burst oscillations. However, burst
oscillations from the accreting millisecond pulsar XTE J1814-338 have
recently led to the first detection of significant harmonics
(Strohmayer et al. \cite{Setal03}). The large number of bursts from
J1814 and the good signal to noise achievable by co-adding bursts
leads to the best chance to date to use burst oscillation signals
detected by RXTE to constrain neutron star parameters. Such work is in
progress at the time of this writing (Bhattacharyya et
al. \cite{Bhatt03b}).

\subsubsection{Pulse Profile Fitting}

As alluded to above, the detailed shapes of rotational modulation
pulses are a unique function of a number of neutron star and binary
orbital parameters. The modulation amplitude depends most sensitively
on the size and shape of the surface emitting area, the viewing
geometry, and the stellar compactness, $\beta$. The sharpness of the
pulse profile (often conveniently expressed in terms of the strength
of harmonics) is dependent on the relativistic beaming introduced by
the surface rotational motion. The maximum observed surface velocity,
$v_{\rm rot}$, is given by $\Omega_{\rm spin} R \sin i$. Since the
spin frequency is known, the velocity is directly related to the
stellar radius and the usually unknown inclination angle, $i$.
Indeed, the pulse profile provides a unique signature of these
different quantities.  That is, if pulse profiles can be measured with
infinite precision, then in principle, a unique solution for the
parameters, $M$, $R$, and $i$ can be obtained. Further, if pulse
profiles can be measured with sufficient precision during the burst
rise, {\it ie. as the X-ray emitting area grows}, then it should also
be possible to see, in snapshot fashion, how the burning spreads
and thus constrain the physics of nuclear flame propagation on neutron
stars. In the next section I will explore the extent to which a future
X-ray timing mission with $\approx 10\times$ the collecting area of
PCA can place constraints on neutron star masses and radii by fitting
the pulse profiles observed during the rise of X-ray bursts.

\begin{figure}
  \includegraphics[height=.3\textheight]{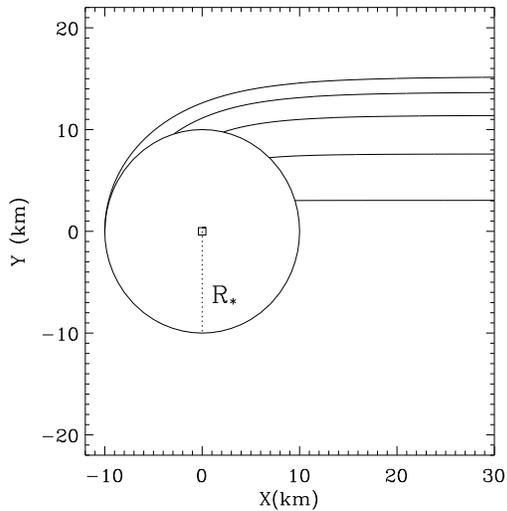}
  \caption{The trajectories of photons from the surface of a neutron
  star with a compactness $\beta = 0.284$. The observer is located at
  $x = + \infty$.}
\end{figure}

\subsection{Spectroscopy: Lines and Continuum}

It has been known for some time that continuum spectral analysis of
Eddington limited X-ray bursts can in principle provide constraints on
neutron star masses and radii (for a brief discussion see the recent
reviews by Lewin, van Paradijs \& Taam \cite{lvpt}; and Strohmayer \&
Bildsten \cite{SB03}).  The method has historically suffered from
several systematic uncertainties; the unknown atmospheric composition,
uncertainties in the intrinsic spectrum (leading to errors in deriving
the effective temperature from the observed color temperature), as
well as uncertainty in what fraction of the neutron star surface is
emitting.  With the discovery of burst oscillations, which provide a
direct indication for asymmetries, this concern has taken on added
importance. Although some of these problems remain, work in this area
with the higher signal to noise RXTE data has continued (see for
example Shaposhnikov, Titarchuk \& Haberl \cite{Shaposh03}).

Perhaps the most direct method of measuring neutron star masses and
radii is by the detection of spectral features (lines and edges)
originating in their surface atmospheres. An observation of an
identified spectral line gives the gravitational redshift, $1 + z =
\left ( 1 - 2\beta \right )^{-1/2}$, at the neutron star surface, 
which provides a direct measurement of the compactness, $\beta$.
Although reliable spectral features from neutron stars have been
notoriously hard to find, recent observations of X-ray bursts from the
LMXB 0748-676 with the XMM/Newton Reflection Grating Spectrometers
have provided evidence for Fe XXVI absorption lines at a redshift of
$z = 0.35$ (see Cottam, Paerels \& Mendez \cite{cottam2002}).

In addition to providing a direct measure of $\beta$, additional mass
- radius information is encoded in the line profile. If the line width
is dominated by rotation of the neutron star, then a measurement of it
constrains the stellar radius through the surface velocity $v_{\rm
rot} = \Omega R \sin i$. For slowly rotating sources, many lines will
be dominated by Stark (pressure) broadening (see Paerels
\cite{Paerels97}), which is proportional to $M/R^2$, so that in either
the rotation or Stark broadening limits, accurate line identifications
and profiles can provide enough information to determine both $M$ and
$R$ uniquely.

For the burst oscillation sources, with known spin frequencies in the
200 - 600 Hz range, rotation should be the dominant broadening
mechanism as long as the system inclination is not too small (see for
example, Ozel \& Psaltis \cite{OP03}). The rotationally dominated line
profiles also contain information on the fraction of the neutron star
surface that is involved in the line formation.  For example, emission
from a fraction of the neutron star surface produces a characteristic
``double-horned'' line profile. Indeed, the relative strengths of the
red and blue wings is sensitive to relativistic gravitational effects,
such as frame dragging (Bhattacharyya \cite{Bhatt03}). The XMM
observations suggest absorption lines with $\approx 10$ eV equivalent
width (Cottam, Paerels \& Mendez \cite{cottam2002}; Bildsten, Chang \&
Paerels \cite{bcp}). In a later section I will explore briefly the
sensitivity of future missions, such as NASA's Constellation-X, to
such lines.

\section{Constraints from Burst Oscillations}

An important capability provided by a larger collecting area is the
increased sensitivity to harmonic content of the pulse profiles.  As
described above this provides information on the surface velocity and
thus the stellar radius for neutron stars with known spin frequencies.
Since both harmonic signals and the pulsed amplitude are strongest
when the emitting area is small, I start by exploring the constraints
that can be made by fitting the observed pulse profiles near the onset
of bursts. 

\subsection{Physics of the Model}

To do this I first generate a model of the time dependent pulse
profile produced by a rotating neutron star.  I build on the modelling
described by Nath, Strohmayer \& Swank \cite{nss02}. The surface
emission is assumed to be a blackbody with temperature, $kT$. The
trajectories of photons emitted from the surface of a neutron star are
in general curved.  These effects are included by assuming that the
external spacetime is the Schwarzschild metric. As an example, Figure
1 shows the paths of photons leaving the surface of a neutron star
with compactness $\beta = 0.284$. Light bending allows a larger
fraction of the neutron star to be seen by a distant observer, and
thus the stronger the light deflection the smaller is the pulsed
amplitude. Relativistic beaming and aberration of the specific
intensity produced by rotation are also included, and arbitrary
viewing geometries are also allowed. For the angular dependence of the
specific intensity I use the limb-darkening law appropriate for a
``grey'' scattering atmosphere (see Chandrasekhar \cite{chandra60}).
The profiles are calculated by summing the contributions from many
small surface area elements on the neutron star surface. Finally, the
spectrum in the rotating frame of the neutron star is appropriately
redshifted to an observer at infinity. This amounts to a total of
eight model parameters, the mass, $M$, the radius, $R$, the orbit
inclination, $i$, the initial angular size of the hot spot,
$\alpha_0$, the angular growth rate of the spot, $v_{\theta}$, the
initial co-lattitude of the spot, $\theta_0$, the surface temperature,
$kT$, and the angular rotation rate of the neutron star,
$\Omega$. This geometry assumes that the rotation axis of the neutron
star is perpendicular to the orbital plane. The level of
sophistication in the model is similar to that employed by several
other researchers, including; Weinberg, Miller \& Lamb \cite{wml01}; Muno,
Ozel \& Chakrabarty \cite{moc02} and Braje, Romani \& Rauch \cite{brr00}.

For the purposes of this work I have specialized to emission from an
expanding, circular hot spot, however, more complex hot spot
geometries can indeed be modelled. Although eight parameters are
required to compute a pulse profile, some of these are known {\it a
priori} or are highly constrained by the observations themselves. The
observed continuum spectrum places good constraints on the surface
temperature, $kT$, and for many anticipated targets the spin frequency
is known. Moreover, the parameters which describe the size and growth
rate of the hot spot are to a large extent constrained by the time
dependence of the phase averaged lightcurve. In terms of the observed
shape of the pulses, the most relevant parameters are the mass,
radius, and initial location of the hot region. Indeed, the most
important degeneracy amongst the parameters results from the fact that
pulsed amplitude, which is a strong function of $\beta$, is also a
strong function of hot spot location.

\begin{figure}
  \includegraphics[height=.3\textheight]{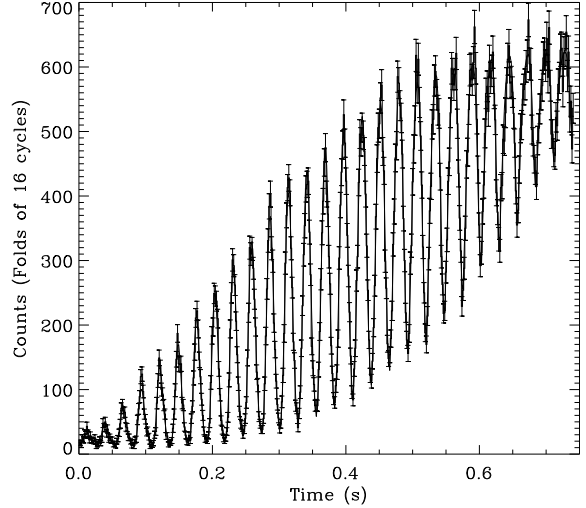}
  \caption{Simulated bolometric pulse profiles for the rising phase of
  an X-ray burst. For this simulation parameters were chosen to mimic
  bursts observed from the LMXB 4U 1636-53. I used $\beta = 0.284$, $R
  = 12$ km, a spin frequency of 582 Hz, and $kT_{\infty} = 2.7$
  keV. Both the observer and the hot spot were located on the
  rotational equator. For clarity each pulse represents a folding of
  16 successive cycles.}
\end{figure}

To obtain a model of the countrate profile seen in a real detector I
take the physical photon spectrum seen at infinity and fold this
through a realistic detector response function. For this I use a
typical RXTE/PCA response matrix, but with the collecting area scaled
up by a factor of 10. This produces a model of the predicted number of
counts seen in the detector as a function of time (rotational
phase). To determine the precision with which parameters can be
estimated, a model profile computed using a set of fixed parameters is
statistically realized a large number of times, and for each
realization the best fitting set of model parameters is found using
$\chi^2$ minimization. As an example, Figure 2 shows a simulated pulse
profile from the model. Figure 3 shows the resulting power density
spectrum computed from this profile, and demonstrates that for this
model, a harmonic signal is easily detected.

\begin{figure}
  \includegraphics[height=.25\textheight]{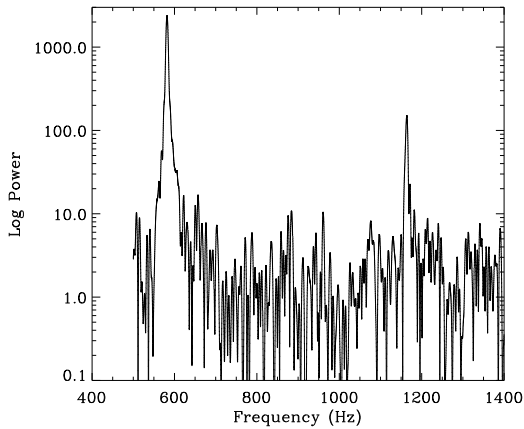}
  \caption{Power density spectrum computed from the lightcurve
  displayed in Figure 2. The vertical axis shows the Log of the Leahy
  normalized power. The fundamental (spin frequency) is at 582 Hz. The
  first harmonic near 1,164 Hz has a Leahy power near 200, and would
  be strongly detected.}
\end{figure}

\subsection{Results}

Due to the large number of parameters and the computational burden
required to compute models, it is time consuming to fully explore the
entire range of the parameter space.  As mentioned above, many of the
parameters are known or well constrained by the continuum spectrum and
phase averaged lightcurve. Therefore, to simplify the problem yet
still retain useful estimates I began by computing models with the hot
spot and viewing geometries fixed while allowing $M$ and $R$ to vary.
The results from this analysis are summarized in Figure 4. I show an
estimate of the confidence region in the mass - radius plane for pulse
profile fits to a single burst from 4U 1636-53 (like that shown in
Figure 2). The small solid contour represents the 1$\sigma$ confidence
region for a ``Super-RXTE'' instrument with $10\times$ the collecting
area of the PCA. For this example the simulations were done for a
neutron star with $\beta = 0.224$, and $R = 12$ km. I also show the
$1\sigma$ contour for a collecting area appropriate to
Constellation-X. As one can see from Figure 4 tight constraints on $M$
and $R$ are in principle achievable, in a statistical sense, with a
factor of 10 increase in collecting area over RXTE/PCA.  Also shown in
Figure 4 are the mass - radius relations for several different neutron
star EOSs. The level of statistical precision would be sufficient to
strongly constrain the mass - radius relation.

\begin{figure}
  \includegraphics[height=.3\textheight]{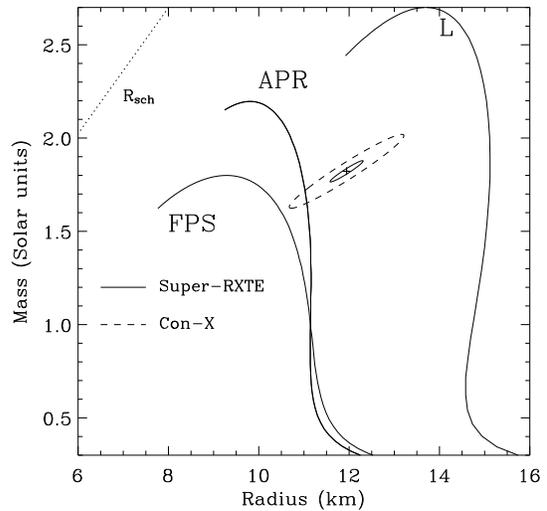}
  \caption{Mass and radius constraints achievable by pulse profile
  fitting with a factor of 10 increase in collecting area over
  RXTE/PCA. The solid and dashed ellipses show, respectively, the
  $1\sigma$ confidence contours for a ``Super-RXTE'' ($10\times$ PCA)
  detector, and for a Constellation-X sized collecting area. The
  results are for pulse fitting during the rise of a single X-ray
  burst using the model described in the text.  The other curves show
  mass - radius relations for several neutron star EOSs; FPS (Lorenz,
  Ravenhall \& Pethick \cite{Lor93}), L (Pandharipande \& Smith
  \cite{PS75}), and APR (Akmal, Pandharipande \& Ravenhall
  \cite{APR98}).}
\end{figure}

\subsection{Caveats: Systematic Uncertainties}

The simulations indicate that on a statistical basis, tight
constraints are achievable, however, one also has to think about
possible systematic uncertainties in the modelling. After all, if the
model is incorrect, then the parameter estimates are also likely to be
incorrect. The two most important sources of systematic errors in the
modelling are likely to be; uncertainties in the neutron star
rest-frame spectrum (including, most importantly, the angular
dependence of the emissivity), and the effects of photon scattering in
the neutron star - accretion disk environment.  In terms of the
spectral modelling the most important quantity with regard to the
pulse profiles is the angular dependence of the emissivity in the
neutron star rest-frame. Miller \& Lamb \cite{ml98} showed that the
angular dependence (in effect, intrinsic beaming of the spectrum) has
an important effect on the amplitude and harmonic content, and thus on
the pulse shape.  If the angular dependence is not correct, then the
mass - radius measurement will have some systematic deviation from the
true values.  It is not immediately obvious how to independently
measure the angular dependence of the emissivity. However, it seems
likely that with sufficient statistical precision the form of the
beaming function can be uniquely constrained by the data, simply
because an incorrect beaming function will not be able to adequately
fit the data, but testing this idea will require more detailed
simulations. It seems likely that we will have to also rely on more
detailed theoretical modelling of the emergent spectrum during
bursts. Fortunately, work in this area is still being done (see for
example Majczyna et al. \cite{Majcy03}; Shaposhnikov, Titarchuk \&
Haberl \cite{Shaposh03}).

We suspect that many X-ray binaries have electron scattering coronae
in their immediate environs.  Scattering of photons emitted by the
neutron star in such a corona can smear out the pulsations, and thus
alter the pulse shapes, this will also introduce a bias in the fitted
parameters. One possible way to mitigate this uncertainty is by
looking at many bursts. It is likely that the properties of the corona
change somewhat with changes in source state and the overall non-burst
spectrum.  If for example the scattering optical depth changes from
burst to burst, then this will cause a movement of the best fit
parameters in the mass - radius plane.  By including a scattering term
in the modelling it may be possible to find the correct scattering
description which brings the fits from many different bursts to a
unique, consistent set of parameters. Again, this could be explored
with more detailed simulations.

\subsection{Theoretical Outlook}

I have presented results based on a relatively simple description of
the relevant physics.  Although this should be sufficient for
exploring the capabilities of future instrumentation, to fully exploit
new observations will require significant theoretical resources.
Fortunately, with the impetus provided by new observations there has
been a corresponding growth in new theoretical thinking about some
long standing issues concerning X-ray bursts.

\begin{itemize}
\item
 Ignition and spreading of nuclear flames on neutron stars. There has
 been great progress on this question in recent years.  In a
 ground-breaking calculation Spitkovsky, Levin \& Ushomirsky
 \cite{SLU02} have shown how rotation of the neutron star is crucial
 in understanding how and where flames ignite and how they spread
 around the star. This and future work opens up the possibility for
 detailed calculations of how the X-ray emitting region grows during a
 burst.
\item
 Nuclear energy release and products of nuclear burning. A number of
 groups have been working on improving models of thermonuclear X-ray
 bursts, both by using more complete nuclear reaction networks and
 more realistic, multi-zone calculations (see for example, Woosley et
 al. \cite{Woos03}; Narayan \& Heyl \cite{nh03}; Schatz, Bildsten \&
 Cumming \cite{Schatz03}). These calculations provide new insight
 into the temperature, flux and composition evolution during bursts.
\item
 Formation of spectra during thermonuclear X-ray bursts. Recent work
 has been done to further explore the angular dependence of the
 emergent spectrum as well as the formation of absorption lines in the
 atmosphere (see Majczyna et al. \cite{Majcy03}; Bildsten, Chang \&
 Paerels \cite{bcp}).
\end{itemize}

These recent calculations represent some of the detailed input physics
required to accurately model the time dependent X-ray flux from a
bursting neutron star. Although a calculation putting all the
different components together has not yet been performed, the required
pieces of the puzzle are largely in place at the present time. It is
not too hard to envision a future effort which attempts to incorporate
these pieces into a unified model which could then be compared with
observations from a future timing mission.

\section{High Resolution Spectroscopy}

A path toward the neutron star EOS which should be less obscured with
possible systematic uncertainties lies in high resolution X-ray
spectroscopy.  As described earlier, accurate line measurements and
resolved profiles (with the correct line identifications) can provide
enough information to obtain both $M$ and $R$ uniquely.  Unfortunately
it has been very difficult to detect any useful lines from neutron
star atmospheres. Recent observations of the thermal emission from
isolated or quiescent neutron stars with the high resolution
capabilities of Chandra and XMM/Newton have been frustratingly devoid
of spectral features (see for example Walter \& Lattimer \cite{WL02};
Burwitz et al. \cite{Burwitz}; Drake et al. \cite{Drake}; Pavlov et
al. \cite{Pavlov}).

A serious problem for non-accreting neutron stars is the high surface
gravity which can sediment out the heavy, line forming metals on a
surprisingly short timescale (see Bildsten, Chang \& Paerels
\cite{bcp}). Because of this, accreting (and thus bursting) neutron stars
may be more promising sources for line detections. An encouraging
recent result is the detection of absorption features in bursts from
the LMXB 0748-676 with XMM/Newton (Cottam, Paerels \& Mendez
\cite{cottam2002}). They found $\approx 10$ eV equivalent width
features by co-adding 28 bursts seen with the RGS spectrometers.
Their proposed identification with Fe XXVI $n=2 - 3$ transitions gives
a redshift of $z = 0.35$ from the neutron star surface.  Although
extremely exciting, this remains so far a single detection, and the
ubiquity of such features in X-ray bursts in general remains to be
established with further observations.

\begin{figure}
  \includegraphics[height=.3\textheight]{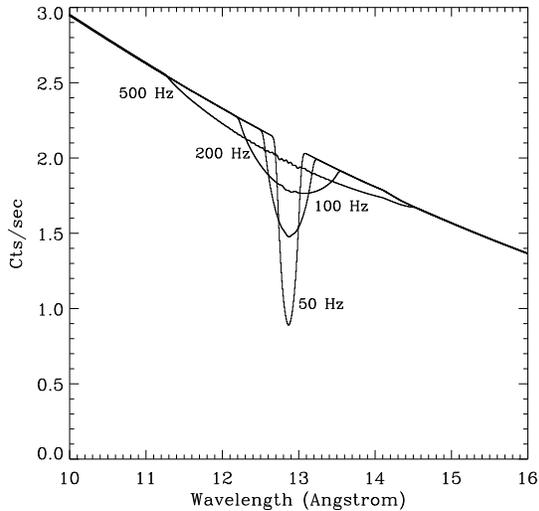}
  \caption{Absorption line profiles from the surface of a rotating
  neutron star. The profiles produced by contributions from the entire
  neutron star surface are shown for several different rotation
  rates. The observer is assumed to be looking in the rotational
  equator.}
\end{figure}

\subsection{Future Capabilities}

A number of future planned missions, such as the Japanese-led AstroE2,
and NASA's Constellation-X, will have high resolution spectroscopic
capabilities. Among these Constellation-X is the most ambitious in
terms of X-ray collecting area, so one can ask the question, will
these missions have the capability to study lines like those seen in
the bursts from EXO 0748-696? To begin to address this question I have
included in the model neutron star spectrum a gaussian absorption line
with a 10 eV equivalent width at the rest energy of the transition
identified by Cottam, Paerels \& Mendez \cite{cottam2002}. I then
calculated the spectrum expected if the star is rotating and the whole
surface is emitting. These spectra were then folded through a
realistic Constellation-X response model to obtain predicted countrate
spectra. As in the timing simulations I normalized the burst flux
using typical bright bursts from the LMXB 4U 1636-53. The results are
rather encouraging. First, Figure 5 shows several line profiles
computed for a range of different rotation rates. For the fastest
rotators, the widths approach $\Delta E/ E \approx 0.1$.  Figure 6
shows a simulation of the countrate spectrum for 15 seconds of
effective exposure in the Constellation-X calorimeter at a flux equal
to the maximum burst flux for 4U 1636-53. I use the flux from 4U
1636-53 as a characteristic value. Although this neutron star spins at
582 Hz, I have used rotation rates of 100 and 200 Hz in these
simulations simply as representative values. Since many bursts last
longer than 15 s, it is not unrealistic to expect that such a spectrum
could be obtained from a single X-ray burst.  The simulation indicates
that in a statistical sense a line feature at this strength can in
principle be detected. Figure 7 shows a second simulation, with a spin
frequency of 200 Hz, and shows the response expected for the
Constellation-X grating as well as the quantum calorimeter. These
simulations, though still simplistic, suggest that high resolution
detectors with large collecting areas will have important
contributions to make for line studies of bursting neutron stars.

\begin{figure}
  \includegraphics[height=.3\textheight]{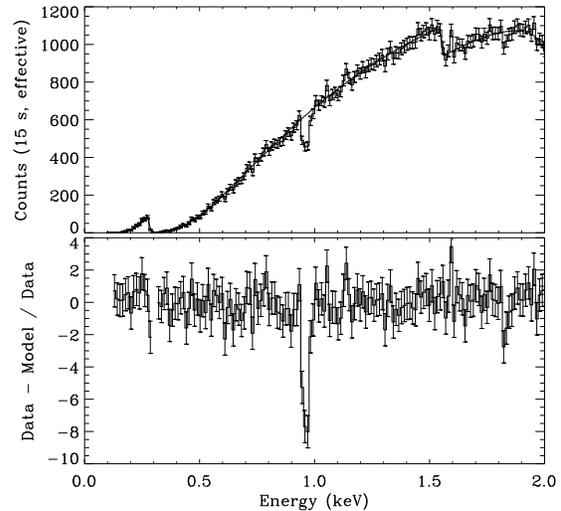}
  \caption{Predicted countrate spectrum (top) and residuals (bottom)
  in the Constellation-X calorimeter for 15 seconds of effective
  exposure at the peak flux of a bright burst from 4U 1636-53. The
  absorption line has an equivalent width of 10 eV, and the rotation
  rate of the neutron star was assumed to be 100 Hz.}
\end{figure}

\begin{figure}
  \includegraphics[height=.3\textheight]{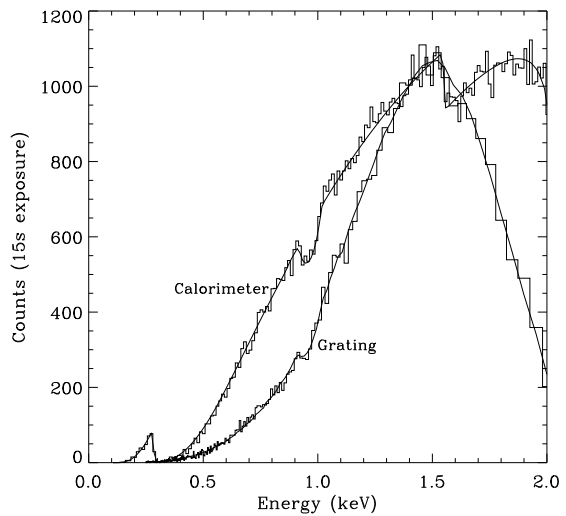}
  \caption{Predicted countrate spectrum in the Constellation-X
  calorimeter and grating for 15 seconds of effective exposure at the
  peak flux of a bright burst from 4U 1636-53. The absorption line has
  an equivalent width of 10 eV, and the rotation rate of the neutron
  star was assumed to be 200 Hz.}
\end{figure}

\section{Summary}

RXTE observations have provided us with several new tools to probe the
interiors of neutron stars. Perhaps the most promising is the detailed
study of burst oscillation pulses, the shapes of which encode
information about the neutron star mass and radius. I have shown that
a future, large area timing mission with about $10\times$ the
collecting area of RXTE/PCA will provide data of sufficient
statistical quality to allow stringent constraints on the neutron star
mass - radius relation and thus the EOS of dense nuclear matter. With
regard to interpretation of the data the primary concern will be the
question of systematic uncertainties associated with the
modeling. There has been substantial advancement in the theoretical
tools needed for such modeling and the pace of these developments
suggests that it is not unrealistic to expect that by the time a
future mission flys, the impact of systematic uncertainties can be
greatly reduced. Moreover, the new data themselves will likely provide
new insights which cannot be anticipated at present.  An additional
strength of attacking the EOS problem using burst oscillations is that
the signals are guaranteed to exist, and it is a virtual certainty
that by studying these oscillations with a factor of 10 better
sensitivity we will learn something new.  

If atmospheric lines are indeed common in bursters, as the new
XMM/Newton results may be indicating, then large area, high resolution
spectroscopy could be the key to unlocking the secrets of the dense
matter EOS.  Current observations do not yet give us enough
information to determine whether the lines are there in sufficient
strength and number to really go after them. It is possible that more
XMM observations will provide the answers. It also seems clear that
Constellation-X will have important contributions to make with regard
to spectral lines from bursters, and if lines are present in sufficient 
strength and number, then stringent EOS constraints may be possible.

\begin{theacknowledgments}

I would like to thank Sudip Bhattacharyya, Cole Miller, Jean Swank,
Craig Markwardt, Will Zhang, Ed Brown, Andrew Cumming, Lars Bildsten,
Mike Muno, Nitya Nath, Jean in't Zand, Erik Kuulkers, Remon
Cornelisse and Anatoly Spitkovsky for sharing various comments,
discussions and ideas related to this work.  I thank the organizers
for the chance to speak at the meeting on this topic.

\end{theacknowledgments}


\bibliographystyle{aipproc}   

\bibliography{strohmayert}

\IfFileExists{\jobname.bbl}{}
 {\typeout{} \typeout{******************************************}
  \typeout{** Please run "bibtex \jobname" to optain} \typeout{** the
  bibliography and then re-run LaTeX} \typeout{** twice to fix the
  references!}  \typeout{******************************************}
  \typeout{} }

\end{document}